\documentclass[aps,prl,twocolumn,superscriptaddress,showpacs]{revtex4}
\usepackage{amssymb}
\usepackage{graphicx}
\usepackage{amsmath}
\usepackage{amsfonts}

\begin{document}

\title{Localized In-Gap States and Quantum Spin Hall Effect in Si-Doped InAs/GaSb Quantum Wells}
\date{\today}

 \author{Dong-Hui Xu}
 \affiliation{Department of Physics, Zhejiang University, Hangzhou 310027, P.R. China}

 \author{Jin-Hua Gao}
 \affiliation{Department of Physics, Huazhong University of Science and Technology, Wuhan, Hubei, China }

 \author{Chao-Xing Liu}
 \affiliation{Department of Physics, The Pennsylvania State University, University Park, Pennsylvania 16802-6300}

 \author{Jin-Hua Sun}
 \affiliation{Department of Physics, Zhejiang University, Hangzhou 310027, P.R. China}

 \author{Fu-Chun Zhang}
 \email{fuchun@hku.hk}
 \affiliation{Department of Physics, and Center of Theoretical and Computational Physics, The University of Hong Kong,
 Hong Kong, China}
 \affiliation{Department of Physics, Zhejiang University, Hangzhou 310027, P.R. China}

 \author{Yi Zhou}
 \email{yizhou@zju.edu.cn}
 \affiliation{Department of Physics, Zhejiang University, Hangzhou 310027, P.R. China}

\begin{abstract}
We study localized in-gap states and quantum spin Hall effect in Si-doped InAs/GaSb quantum wells.
We propose a model with donor and/or acceptor impurities to describe Si dopants.
This model shows in-gap bound states and wide conductance plateau with the quantized value $2e^2/h$ in light dopant concentration,
consistent with recent experiments by Du et al.\cite{Du13} We predict a conductance dip structure due to backward scattering
in the region where the localization length $\xi$ is comparable with the sample width $L_y$ but much smaller than the sample length $L_x$.
\end{abstract}










\maketitle

The quantum spin Hall (QSH) insulator is the first theoretically predicted two-dimensional (2D)
time reversal invariant topological insulator (TI) that manifests topologically
non-trivial edge states. It is characterized by topologically robust
gapless (spin polarized) counter-propagating states at the edge while there
exists an energy gap in the bulk.\cite{KaneMele,Bernevig2006} The first realistic
material for QSH insulators was proposed by Bernevig \textit{et al.}\cite{BHZ} in a semiconductor quantum well (QW)
formed by a HgTe layer sandwiched between two CdTe layers when the HgTe layer
exceeds a critical width, $d_c \sim 6.3$nm. The edge transport
channels in HgTe QWs were later discovered by K\"{o}nig \textit{et al.} in transport experiments,\cite{Molenkamp} which confirmed theoretical predictions
of edge states in TIs and opened up the experimental investigation of the
QSH insulators and other TIs.\cite{HasanRMP,QiRMP,moore} Although non-local measurements
have exhibited edge transports,\cite{Roth} large conductance
fluctuation in the QSH regime of HgTe QWs has never been well understood.

The second example of QSH insulators is the type II InAs/GaSb QW proposed by Liu \textit{%
et al.}\cite{cxliu} and has been realized in experiments \cite%
{Du11,Du12,Suzuki13}. In InAs/GaSb QWs, the conduction band of InAs is about 150 meV lower than the
valence band of GaSb (see Fig.1(a)), forming a so-called ``inverted band structure''. The position of electron and hole
subbands in this system can be controlled by varying the thickness of InAs
and GaSb layers and the band inversion occurs in a large range of the thickness of the QW.
The hybridization of electron and hole states in such a QW
opens a mini-gap of 40$\sim$60K. When the Fermi energy lies in the mini-gap,
charge transport is dominated by topological edge modes, which can be extracted from conductance measurements.\cite{Du11,Du12}
However, early experiments always exhibit conductance larger than the expected values of $\frac{2e^2}{h}$ for a QSH insulator, 
presumably due to residual bulk currents carried by disorder induced (extended) states inside the mini-gap.

\begin{figure}[htbp]
\centering \includegraphics[width=8.4cm]{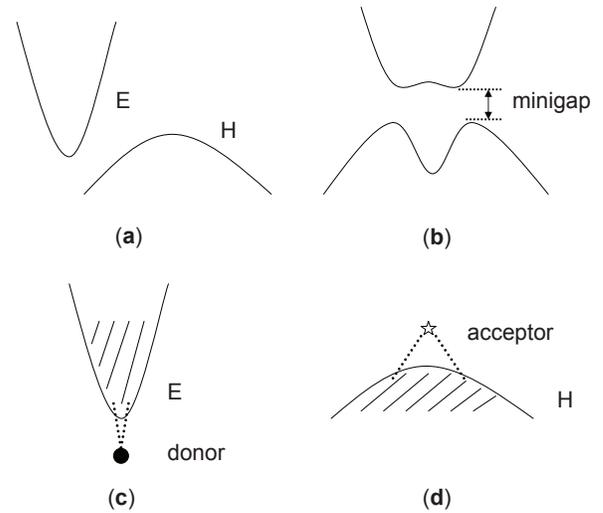}
\caption{(a) and (b) Band inversion in InAs/GaSb. (a) Conduction band of
InAs (denoted by $E$) is about 150 meV lower than the valence band of GaSb
(denoted by $H$). (b) The hybridization of the electron and hole states
opens a mini-gap. (c)-(d) Schematic impurity level and impurity scattering.
(c) Si atoms serve as donors in InAs. (d) Si atoms serve as acceptors in
GaSb. }
\end{figure}

Very recently, Du et al. \cite{Du13} introduces a small amount of Si dopants (%
$\sim 10^{11}$cm$^{-2}$) in InAs/GaSb QWs. Si atoms may serve
as donors in InAs and acceptors in GaSb. The bulk conductance is found to be suppressed by Si doping,
resulting in a mobility gap of 26K inside the mini-gap. Consequently,
charge transport can only be carried by edge currents, leading to the observation of
the quantized conductance  $G_0=2e^2/h$ in experiments. More fascinatingly,
as a function of the front gate voltage, wide conductance plateaus are observed with
a small conductance fluctuation within $1\%$ of the quantized value.
Therefore, it is natural to ask how the Si-doping induces the quantized conductance plateau. 

Theoretically, the disorder effect has been numerically investigated in the context of graphene\cite{Nagaosa07} and
HgTe QWs\cite{joseph,Shen09}. It was found that under magnetic fields,
disorder can induce localization behavior of helical edge states,
which is the origin of the cusp-like feature in magneto-transport of HgTe QWs.\cite{joseph}
In addition, strong disorder may drive an
ordinary insulating state to a topologically non-trivial state with a quantized conductance of $G_0$.\cite{Shen09}
The latter is called the topological Anderson insulator (TAI),
where the disorder is also responsible for generating extended edge states. The
existence of TAI was confirmed in independent numerical simulations\cite%
{Xie09} and may be understood as band inversion caused by effective mass
renormalization\cite{Beenakker09}. 
(Energy band renormalization is distinct in HgTe and InAs/GaSb, see Supplementary Materials for details.)
The phenomenon of TAI requires very strong impurity scattering potential or high impurity concentration
to localize the bulk states and to renormalize the effective mass.
However, all these studies have not paid close attention to the role of in-gap states
as observed in the Si-doped InAs/GaSb QWs.
This motivates us to critically consider the effect of Si-dopant.

In this Letter, we study the effect of Si-dopants in InAs/GaSb QW in
the band inverted region. A single Si-dopant serves as a donor to conduction
band or an acceptor to valence band and introduces a bound state in the mini-gap,
similar to a charge-impurity bound state in conventional semiconductors.
Note that weak disorder leads to localization in 2D in the thermodynamic limit. For a mesoscopic sample of length $L_x$, 
the nature of localization depends on the localization length $\xi$.
When $\xi< L_x$ in QWs, a bulk mobility gap is opened and only the edge state transport remains.
We use Landauer-B\"{u}ttiker formalism to calculate conductance and show a wide plateau
in quantized conductance of $2e^2/h$. Our theory explains the observed mobility gap
and quantized conductance in the experiment of Du et al.

\textit{Model Hamiltonian of InAs/GaSb QWs:} The bulk of InAs/GaSb QW can be well described by a
four band tight-binding model on a square lattice where four relevant atomic
states $\left\{ \left\vert E+\right\rangle ,\left\vert E-\right\rangle
,\left\vert H+\right\rangle ,\left\vert H-\right\rangle \right\} $ are
involved, where $E$ and $H$ mark electron and hole states respectively and $%
\pm $ correspond to pseudo-spins. This tight-binding Hamiltonian can be
derived from a corresponding $k\cdot p$ Hamiltonian\cite{cxliu} and reads
\begin{equation}
H_{0}=\sum_{i\sigma \alpha }V_{\alpha \sigma }c_{i\alpha \sigma }^{\dagger
}c_{i\alpha \sigma }+\sum_{i\tau \sigma \alpha \beta }t_{\alpha \beta
}^{\tau \sigma }c_{i\alpha \sigma }^{\dagger }c_{i+\tau \beta \sigma },
\label{H0}
\end{equation}%
where $i$ is the site labeling, $\tau =\pm \hat{x},\pm \hat{y}$ denotes the
four nearest neighbors bond, $\sigma =\pm $ is for pseudo-spin, and $\alpha
,\beta =E,H$ is the orbital index. The pseudo-spin is a good quantum number.
In the sub-Hilbert space spanned by $\left\{ \left\vert E \sigma \right\rangle ,\left\vert H \sigma \right\rangle \right\} $,
$V_{\alpha \sigma }$ is a diagonal matrix
\begin{equation}
V_{\sigma }=\left(
\begin{array}{cc}
C-4D+M-4B & 0 \\
0 & C-4D-M+4B%
\end{array}%
\right)  \label{V0}
\end{equation}%
and $t_{\alpha \beta }^{\tau \sigma }$ is given by the following matrix form%
\begin{eqnarray}
t^{\pm \hat{x}\sigma } &=&\left(
\begin{array}{cc}
D+B & \mp i\sigma A/2 \\
\mp i\sigma A/2 & D-B%
\end{array}%
\right) ,  \notag \\
t^{\pm \hat{y}\sigma } &=&\left(
\begin{array}{cc}
D+B & \pm A/2 \\
\mp A/2 & D-B%
\end{array}%
\right) ,  \label{t0}
\end{eqnarray}%
where $A$, $B$, $C$, $D$ and $M$ are parameters which determine the band
structure and will be given later. The lattice constant in InAs/GaSb is
about $6$\AA . But we can choose a different lattice constant in the
tight-binding model by properly choosing parameters $A$, $B$, $C$, $D$ and $M $,
since it is an effective model derived from the $k\cdot p$ Hamiltonian.
In this letter, we shall set the lattice constant as $a=20$\AA\ and choose
$A=0.0185$eV, $B=-0.165$eV, $C=0$, $D=-0.0145$eV and $M=-0.0078$eV. This set
of parameters corresponds to the set of $k\cdot p$ parameters used in Ref. \cite{cxliu}.
For simplicity, we neglect the terms describing bulk inversion asymmetry and structure inversion asymmetry,\cite{cxliu}
which are inessential for the physics of QSH effect.

By Fourier transformation, the $\mathrm{k}$-component of Hamiltonian (\ref%
{H0}) can be expressed as the following $4\times 4$ matrix
\begin{equation}
H_{0}(\mathrm{k})=\epsilon (\mathrm{k})\mathbf{I}+\left(
\begin{array}{cc}
h_{0}(\mathrm{k}) & 0 \\
0 & h_{0}^{\ast }(-\mathrm{k})%
\end{array}%
\right) ,  \label{H0k}
\end{equation}%
where $h_{0}(\mathrm{k})$ is a $2\times 2$ matrix,
\begin{equation}
h_{0}(\mathrm{k})=\left(
\begin{array}{cc}
M(\mathrm{k}) & A(\sin k_{x}-i\sin k_{y}) \\
A(\sin k_{x}+i\sin k_{y}) & -M(\mathrm{k})%
\end{array}%
\right) ,  \label{h0k}
\end{equation}%
$\mathbf{I}$ is the $4\times 4$ identity matrix, $M(\mathrm{k})=M-2B(2-\cos
k_{x}-\cos k_{y})$ and $\epsilon (\mathrm{k})=C-2D(2-\cos k_{x}-\cos k_{y})$%
. Eq. (\ref{H0k}) recovers the $k\cdot p$ Hamiltonian at small $k$\cite%
{cxliu}. According to Fu and Kane's criterion for TIs, topological
nontrivial states exist when $0<M/2B<2$.\cite{FuKane07}

Now we consider the disorder induced by Si dopants, which serve
as donors in InAs ($E$ band) and acceptors in GaSb ($H$ band).
Donors and acceptors in semiconductors can be treated as hydrogenic ions
with positive or negative charges respectively.
A donor attracts $E$ band electrons and its interaction with $H$ band can be
neglected. Similarly, an acceptor provides a repulsive scattering potential to $H$
band while leaves $E$ band non-interacted.\cite{semiconductor} To model the disorder
effect of Si dopants in InAs/GaSb, we introduce the following on-site
impurity Hamiltonian,
\begin{equation}
H_{imp}=-\sum_{i\in R_{d},\sigma }V_{i}c_{i,E\sigma }^{\dagger }c_{i,E\sigma
}+\sum_{i\in R_{a},\sigma }V_{i}c_{i,H\sigma }^{\dagger }c_{i,H\sigma },
\label{Himp}
\end{equation}%
where $R_{d}$ is the collect of site $i$ where the Si dopant serves as
donors and $R_{a}$ denotes the site set for acceptors. $V_{i}$ is positive
and distributes randomly within a range of $(V_{\text{min}},V_{\text{max}})$.
While this is a simplified model for the charge impurity potentials, the model captures the basic physics in Si-dopant systems. As we will see below, this type of disorder has the largest efficiency in inducing in-gap bound states,
which are essential for the localization physics.
For the lattice constant $a=20$\AA\ used in this Letter, $1\%$ impurity in
our model corresponds to $2.5\times 10^{11}$cm$^{-2}$ Si dopants, and $%
10^{11}$cm$^{-2}$ Si dopant corresponds to $0.4\%$ impurity in our model.

\begin{figure}[htbp]
\centering \includegraphics[width=8.4cm]{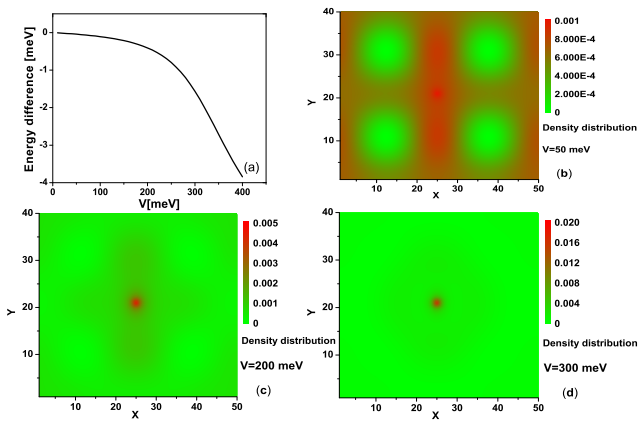}
\caption{(Color online) (a) Binding energy of single impurity (donor) state. (b)-(d)
Local density of states of bound states induced by a single impurity.
$V=50,200,300$meV in (b)-(d). }
\label{ldos}
\end{figure}

\begin{figure}[htbp]
\centering \includegraphics[width=8.4cm]{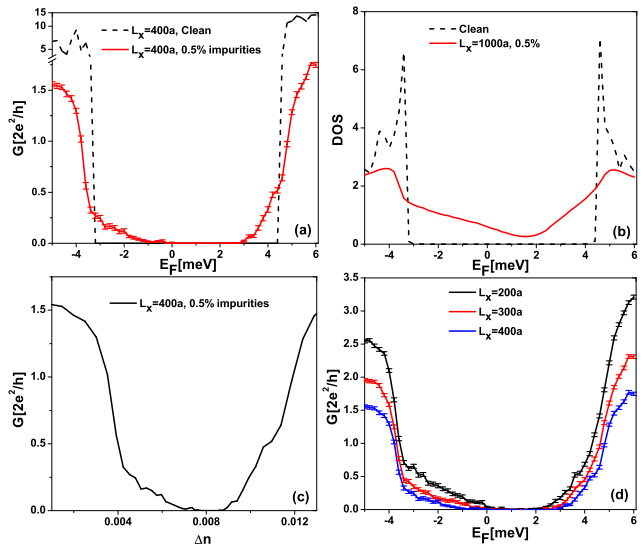}
\caption{(Color online) Conductance and DOS in the presence of impurities.
The width of strip is set to $L_y=200a$ and the lattice constant $a=2$nm. The
 concentration of impurity is $0.5 \% $.
(a) Conductance for clean sample and impure sample with length $L_x=400a$.
(b) Lead-free DOS for clean sample and impure sample with length $L_x=1000a$.
(c) Conductance as the function of electron density with length $L_x=400a$.
(d) Conductance of different samples with lengths $L_x=400a$ (blue), $300a$(red) and $200a$(black).}
\label{bulk}
\end{figure}

\textit{Single Impurity:} We begin with a single impurity problem.
In a semiconductor, a dopant can always induce in-gap bound states because of
the long ranged Coulomb attraction to conduction band electrons or valence band holes.
In our model, the impurity potential is a short ranged and delta function like potential in 2D.
At first glance, this kind of impurity potential does not support bound states unless
it exceeds a threshold. But this is not true here because the density of states (DOS) is singular
at band edge in the clean system (see Fig.\ref{bulk}(b)). In this case, even infinitesimal
attractive potential will give rise to in-gap bound states\cite{TKNg}.
So our model catches the in-gap physics correctly although it uses
a simplified impurity potential.
To examine this point, we diagonalize the Hamiltonian $H=H_{0}+H_{imp}$ numerically
on a $50\times 40$ lattice, and find that an in-gap state appears in the presence
of a single impurity. For simplicity, we only show the results on donors, and
the acceptor situation is similar.
The energy difference $\Delta E$ between the in-gap state and
band edge for a pure system is plotted in Fig.\ref{ldos}(a) as a function of
the attractive potential $V$. It is clear that there exist bound states
at any attraction strength.
The density distributions for the corresponding in-gap bound
states are shown in Fig.\ref{ldos}(b)-(d). The in-gap state becomes more and more localized
as the attractive potential increases.
For hole states in the valence band, an acceptor carries a negative charge
and is attractive to holes while repulsive to electrons. It will
also induce in-gap bound states. Therefore, our choice of signs of $V_i$ in $H_{imp}$
in Eq.(\ref{Himp}) always gives rise to in-gap bound states.

\textit{Localization in bulk:} We continue to study the localized states in
bulk in the presence of many Si dopants. We consider half of the Si dopants are donors
and the other half are acceptors. Hereafter we set $V_{\text{min}}=200$meV
and $V_{\text{max}}=300$meV unless specified otherwise. To investigate the
localization problem, we calculate DOS and conductance $G=G_{xx}$
of the Hamiltonian $H=H_{0}+H_{imp}$ by using the recursive Green's
function method \cite{rgf} and Landauer-B\"{u}ttiker-Fisher-Lee formula \cite{LB}.
We consider a strip geometry consisting of a
rectangular disorder region (with length $L_{x}$ and width $L_{y}$) and two
semi-infinite doped metallic leads connected to the rectangle along the $x$ direction.
This setup allows us to study the transport coefficients with both open boundary condition
(OBC) and periodic boundary condition (PBC) along the $y$ direction.
We shall utilize PBC to study the bulk states and OBC to study the edge states.

The results are shown in Fig. \ref{bulk}. There is a window in the Fermi energy, $E_F\sim 0.2-2.4$meV,
where DOS is finite but the conductance vanishes. This means that there exist fully localized in-gap states
with the localization length much smaller than $L_x$ in this regime.
For the Fermi energy in the regime $-3.0$meV$<E_F<0.2$meV or $2.4$meV$<E_F<4.0$meV, although the Fermi energy
is still in the bulk gap, the conductance is non-zero. This is due to the finite size effect.
Namely, when the localization length $\xi$ of localized states is comparable with or larger than the system size $L_x$,
the size effect become considerable and bulk transport is allowed in this case. In Fig. \ref{bulk}(d),
we increase the length $L_x$ and the regime for zero conductance is also enlarged.

The localized in-gap states do not contribute to transport thereby the conductance.
On the other hand, these states contribute to the total density of states revealed in capacity measurements.
The existence of large amount of localized in-gap states in DOS in our simulation qualitatively agrees
with the capacity measurement in experiments\cite{Du13}.

We remark that the sharp impurity potential scenario we use for Si-dopant is very different from
the weak localization scenario. Firstly, increasing impurity concentration will
result in a finite bulk conductance due to the formation of impurity bands. Secondly, smooth disordered potential
may give rise to in-gap ``extended" states with a long localization length $\xi\gg L_x$ due to large density of state at the band edge. Correspondingly, the mobility gap becomes smaller than the mini-gap in the present system. These extended states may be further localized by
Si-dopants described by a sharp impurity potential. (See supplementary materials for details.) This may explain the experiment by Du et al., where the bulk conductance is finite in the absence of the Si-doping and vanishes as Si-impurities are introduced. 

\begin{figure}[htbp]
\centering \includegraphics[width=8.4cm]{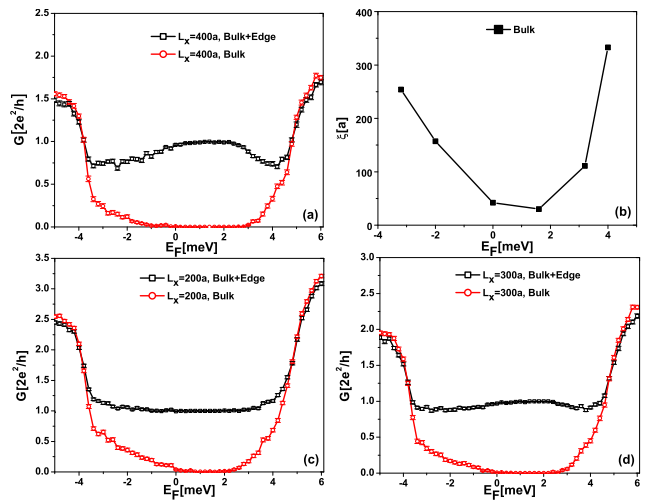}
\caption{(Color online) Quantized conductance plateau and localization length in a strip with width $L_y=200a$ and $0.5\%$ impurity of half donors
and half acceptors.
(a) Conductance in a $400\times 200$ sample.
(b) Localization length as function of Fermi energy.
(c) Conductance in a $200\times 200$ sample.
(d) Conductance in a $300\times 200$ sample. }
\label{edge}
\end{figure}

{\it Edge transport:} When the bulk states are localized by impurities,
i.e.,  $\xi \ll L_x$,
the transport is entirely dominated by the edge channels, as clearly seen in Fig. \ref{edge}(a).
In the regime $0.2$meV$<E_F<2.4$meV, a wide conductance plateau with the quantized value
$\frac{2e^2}{h}$ emerges for OBC while bulk conductance is zero and all bulk states
are localized for PBC. In this regime, the localization length,
$\xi\equiv-2\lim_{L_{x}\rightarrow\infty}L_{x}\langle\ln G/G_{0}\rangle^{-1}$ \cite{RMT},
is found to be much smaller than both the length $L_x$
and width $L_y$ of the system, as shown in Fig. \ref{edge}(b).
To estimate the localization length $\xi$, we use PBC along the $y$ direction and scale
the bulk conductance $G$ with varying length $L_{x}$ and fixed width $L_{y}$ at samples.

When the Fermi energy is tuned to the regime $-3$meV$<E_F<0.2$meV or $2.4$meV$<E_F<4$meV,
we find that although the bulk conductance increases according to the calculation with PBC,
the total conductance for a system computed with OBC decreases at plateau edge.
This conductance dip is a finite size effect. In this regime, the localization length $\xi$
increases rapidly and when it is comparable to the sample width but much smaller than the sample length ($L_y\sim\xi\ll L_x$),
the electron at one edge state can interfere with in-gap bound states in the bulk
and will be scattered to the opposite edge. This will lead to a significant reduction
of edge currents. We emphasize here that the penetration length of the edge states in a clean system
is quite small for the present parameters, which can not induce direct hybridization between
two edge states. The backscattering is mediated by in-gap bound states.


For a given impurity concentration, the dip structure in conductance may smooth out by
increasing sample width $L_y$ or decreasing sample length $L_x$. Because the former will
reduce the coupling between edge states and in-gap bulk states, while the later will increase
the bulk conductance. Figs. \ref{edge}(c), (d) and (a) show how
this dip structure emerges with inceasing $L_x$ and fixed $L_y\sim \xi$. This predicted dip structure
in conductance under the condition $L_y\sim\xi\ll L_x$ may be examined in future experiments.

{\it Comparison with experiments and summary:}
Now we shall relate our numerical simulation on the localization effect to the recent experiment of InAs/GaSb QWs with Si doping.
Experimentally, the conductance quantization is only observed at low temperatures.
The localization length $\xi$ is temperature dependent. At high temperature, $\xi$ is
much longer than the sample length $L_x$, the impurity state is extended,
and conductance is higher and not quantized.  As temperature is lowered, $\xi$ of the states
deep in the band gap becomes shorter than the sample size. As a result, the system starts to show insulating behavior
when the Fermi energy is tuned deep into mini-gap. This corresponds to the development of
the mobility gap observed by both the bulk conductance measurement in a Carbino disk geometry and
the electric capacity measurement. In this regime, for a finite sample,
a robust quantized conductance plateau appears due to the edge transport and the system enters into the QSH regime.

Finally, we would like to emphasize the uniqueness of disorder effect in InAs/GaSb QWs. 
In conventional materials, even smooth disorder could lead to a large mobility gap. 
Therefore, the precise quantization conductance plateaus can be easily observed in the quantum Hall effect.\cite{Huckestein} 
In contrast, our numerical calculations have shown that the mobility gap strongly depends on the strength 
and types of disorders in InAs/GaSb QWs due to the singularity of density of states at the band edge. 
Thus, a model with lightly Si doping is required to introduce in-gap bound states to form a mobility gap at a low temperature. 
Our theory explains the essential role of in-gap states for the highly quantized conductance plateau observed in Du et al.'s experiment
at zero magnetic field, and examined conditions for the accurate quantization in the quantum spin Hall systems. 
We note that the experiments of Du et al. also report the extreme robustness of the quantization against external magnetic field, 
which remains to be a theoretical challenge for our future work.

We thank R. Du for many insightful discussions on their experiments, and S. Q. Shen and M. Ma for discussions in the early stage of the project.
This work is partially supported by National Basic Research Program of China
(No.2011CBA00103/2014CB921201), NSFC (No.11374256/11274269), the
Fundamental Research Funds for the Central Universities in China, and HK RGC/GRF grant No. 701010.

\begin{center}
{\bf Supplementary Materials for ``Localized In-Gap States and Quantum Spin Hall Effect in Si-Doped InAs/GaSb Quantum Wells"}
\end{center}

\section{Band renormalization by Anderson disorder: self-consistent Born approximation}

To compare different band renormalization effect by Anderson disorder between HgTe and InAs/GaSb QWs,
we use self-consistent Born approximation to investigate the following random on-site potential,
\begin{equation*}
H_{\textrm{I}}=\sum_{i\sigma \alpha }V_{i}c_{i\alpha \sigma }^{\dagger
}c_{i\alpha \sigma }
\end{equation*}
where $V\in[-W/2,W/2]$ with disorder strength $W$. This type of impurities will renormalize the
energy band through the self-energy, which is defined as
\begin{equation*}
(E_{\textrm{F}}-h_{0}-\Sigma)^{-1}=\langle(E_{\textrm{F}}-h)^{-1}\rangle,
\end{equation*}
where $\langle\cdot\cdot\cdot\rangle$ denotes the disorder average. Here $\Sigma$ is a $2\times2$
matrix which can be decomposed into Pauli matrices: $\Sigma=\Sigma_{\mu}\sigma_{\mu}
\;\;\mu=0,1,2,3.$  The renormalized topological mass and chemical potential are then
given by
\begin{equation*}
\bar{M}=M+\textrm{Re}\Sigma_{3}, \;\; \bar{E_{\textrm{F}}}=E_{\textrm{F}}-\textrm{Re}\Sigma_{0},
\end{equation*}
where $M$ and $E_F$ are bare mass and chemical potential respectively.

In the self-consistent Born approximation, we keep the self-energy up to the second order of $W$ in the spirit of perturbation.
The self-energy is given by
\begin{equation*}
\Sigma=\frac{W^2}{12}\sum_{\mathbf{k}}[E_{\textrm{F}}+i0^{+}-h_{0}(\mathbf{k})-\Sigma]^{-1}.
\end{equation*}
The band gap $E_{gap}$ is determined by the upper band edge $E_u$ and the lower band edge $E_l$ through $E_{gap}=E_u-E_l$,
where $E_u$ reads
\[
E_{u}=\min\biggl[\epsilon(k)+\textrm{Re}\Sigma_{0} +\sqrt{\bar{M}^2(k)+A^2\sin^2k}\biggr],
\]
and $E_l$ reads
\[
E_{l}=\max\biggl[\epsilon(k)+\textrm{Re}\Sigma_{0}-\sqrt{\bar{M}^2(k)+A^2\sin^2k}\biggr],
\]
with $\bar{M}(k)=\bar{M}-2B(2-\cos k_{x}-\cos k_{y})$.

\begin{figure}[h]
\renewcommand\thefigure{S1}
\centering \includegraphics[width=6.4cm]{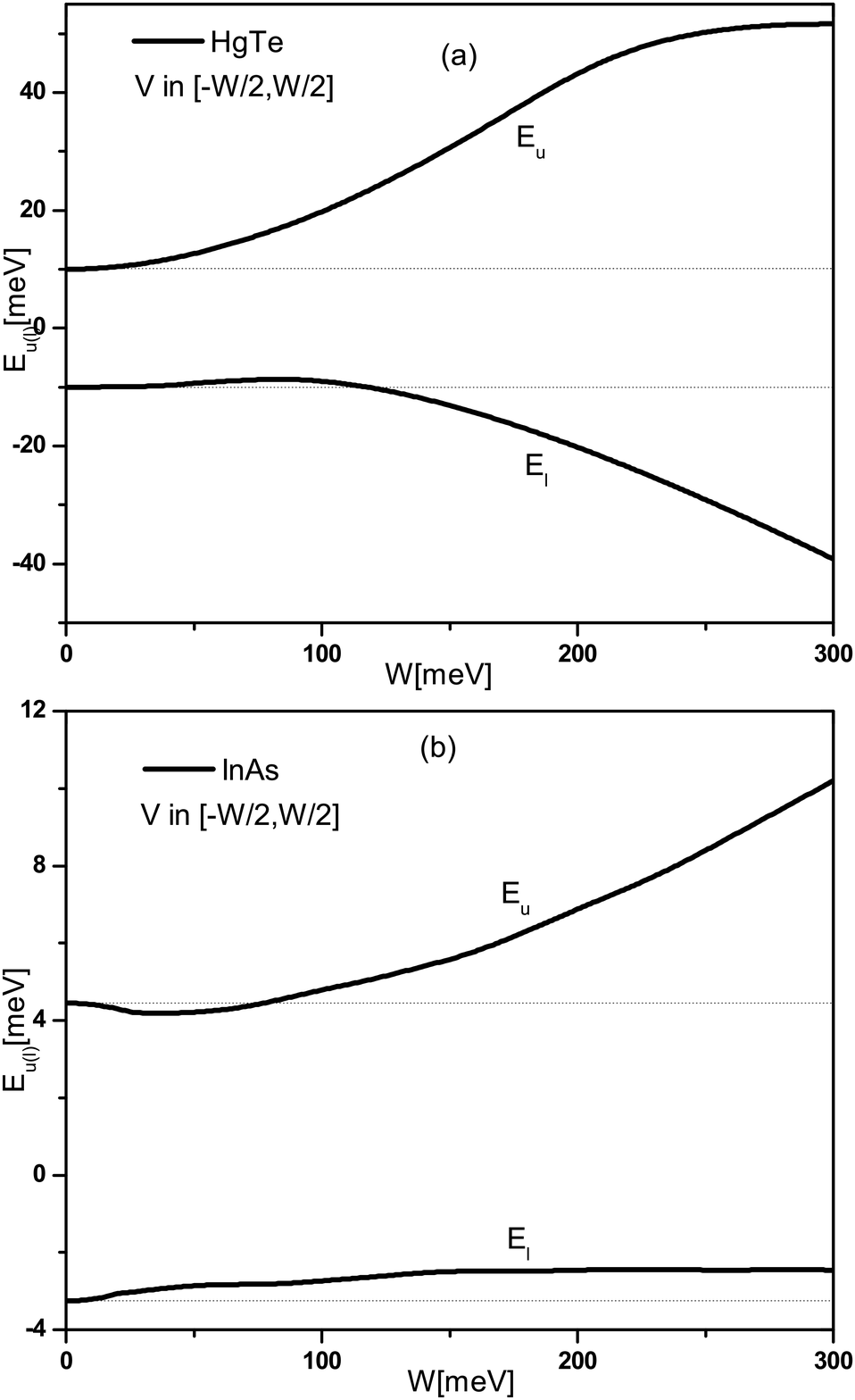}
\caption{Energy band renormalization by Anderson disorder.
(a) Upper and lower band edges in HgTe QW.
(b) Upper and lower band edges in InAs/GaSb QW.
}
\label{scba}
\end{figure}

To compare HgTe with InAs/GaSb, we choose the following bare band parameters for HgTe QW,
$A=0.0729$eV, $B=-0.02744$eV, $C=0$, $D=-0.02048$eV and $M=-0.01$eV,
while use the same paramters as in the main text for InAs/GaSb QW, namely,
$A=0.0185$eV, $B=-0.165$eV, $C=0$, $D=-0.0145$eV and $M=-0.0078$eV.
Fig. \ref{scba} shows the renormalized band edges by disorder
average based on self-consistent Born approximation. One sees that the band gap in InAs/GaSb QW will be narrowed in weak
disorder strength in contrast to HgTe QW where the band gap is enlarged by disorder. 
This indicates that it is easier to induce in-gap states by disorder in InAs/GaSb than that in HgTe.

\section{The formation of impurity band}

\begin{figure}[htbp]
\renewcommand\thefigure{S2}
\centering \includegraphics[width=8.0cm]{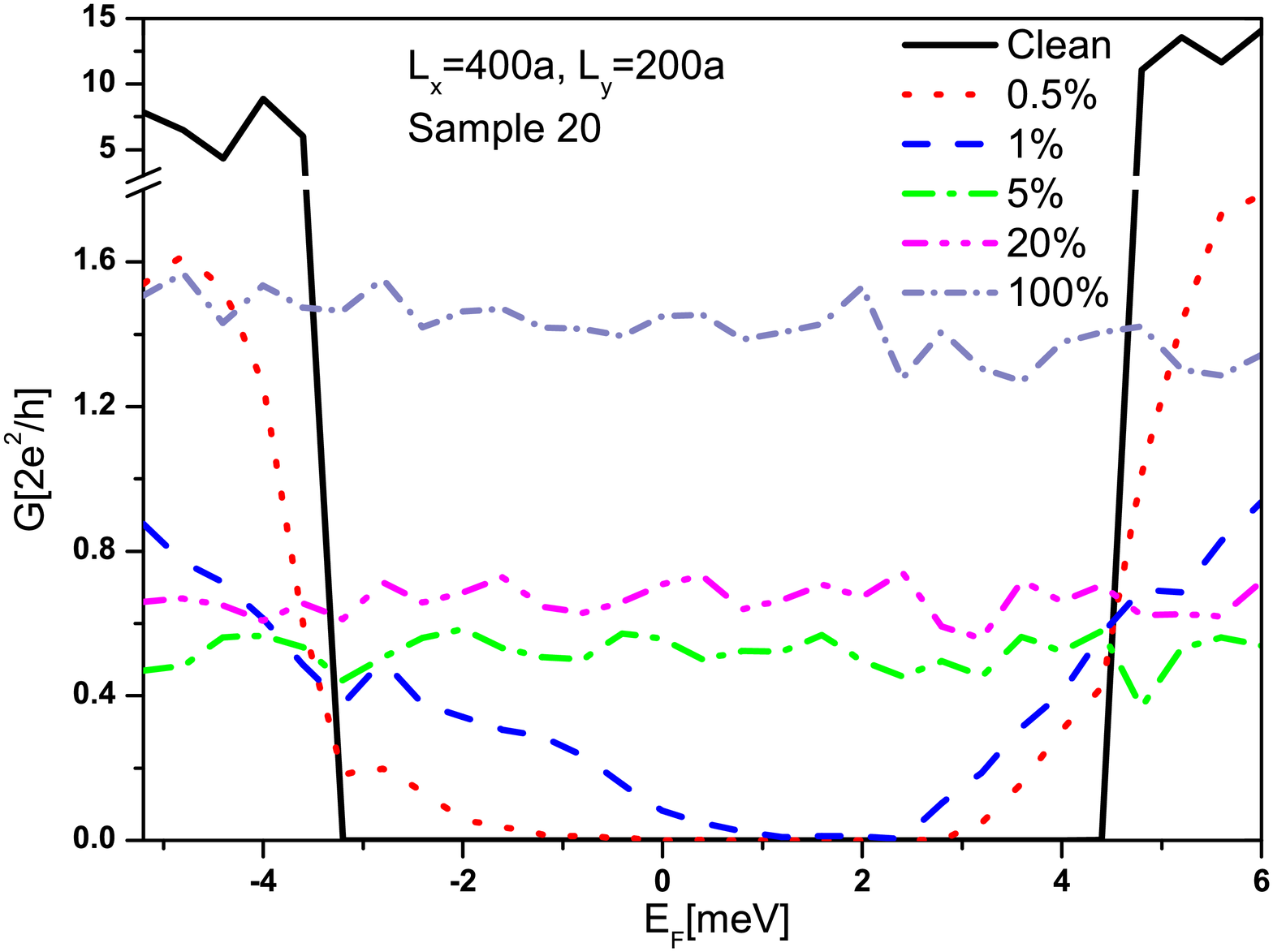}
\caption{(Color online) Bulk conductance changes with the impurity concentration.
In the presence of dilute impurities, bulk conductance vanishes due to strong localization.
As impurity concentration increases, wave function overlap between neigboring
localized states will increase, resulting in finite bulk conductance.
}
\label{impurity_band}
\end{figure}

In the weak localization picture, the localization length will decrease as impurity concentration increases.
In contrast, localized states caused by the strong localization mechanism proposed in the main text will become
more and more ``extended" when impurity concentration increases, because the wave function overlap between
neigboring localized states will increase. Thus the localized states induced by individual impurity will form an impurity
band when the impurity concentration is large enough. This impurity band will contribute to bulk transport, resulting in
finte bulk conductance as shown in Fig.\ref{impurity_band}.

\section{Localization of in-gap ``extended" states induced by smooth disorders}

In addition to sharp impurity potential induced by dilute Si dopants, one may consider smooth
disorder potential commonly used in quantum Hall and other systems.
We find that smooth disorders may introduce in-gap ``extended" states, or to be precise,
localized states with localization length $\xi$ much longer than sample length $L_x$. Moreover, these
in-gap ``extended" states will be localized by sharp impurity potential induced by dilute Si dopants.

To illustrate these effects, we use a random Gaussian potentials $H_{G}$ to model smooth disorder,
\begin{equation*}
H_{G}=\sum_{i,\sigma }V_{G}(i)(c_{i,E\sigma }^{\dagger }c_{i,E\sigma
}+c_{i,H\sigma }^{\dagger }c_{i,H\sigma }),
\label{HG}
\end{equation*}
with
\begin{equation*}
V_{G}(i)=\sum_{j\in R_{G}}V_{0}(j)\exp[-\frac{(\vec{r}_{i}-\vec{r}_{j})^2}{2b^2}],
\label{VG}
\end{equation*}
where $R_{G}$ is the collect of site $j$ where the Gaussian peaks locate, $b$ is a constant to describe
the width of Gaussian potential, $V_0$ distributes uniform randomly in $[-V_{s},V_{s}]$.
We choose $R_G$ to contain $10\%$ sites in the whole lattice, $b=5a$ and $V_s=10$meV to generate smooth
disorder potential.
The sharp impurity potential induced by Si is described by random delta-function-like potential $H_{imp}$
as in the main text with $V_{\text{min}}=300$meV and $V_{\text{max}}=400$meV. We use $4\%$ sharp impurity
concentration to illustrate the localization effect.

\begin{figure}[htbp]
\renewcommand\thefigure{S3}
\centering \includegraphics[width=8.0cm]{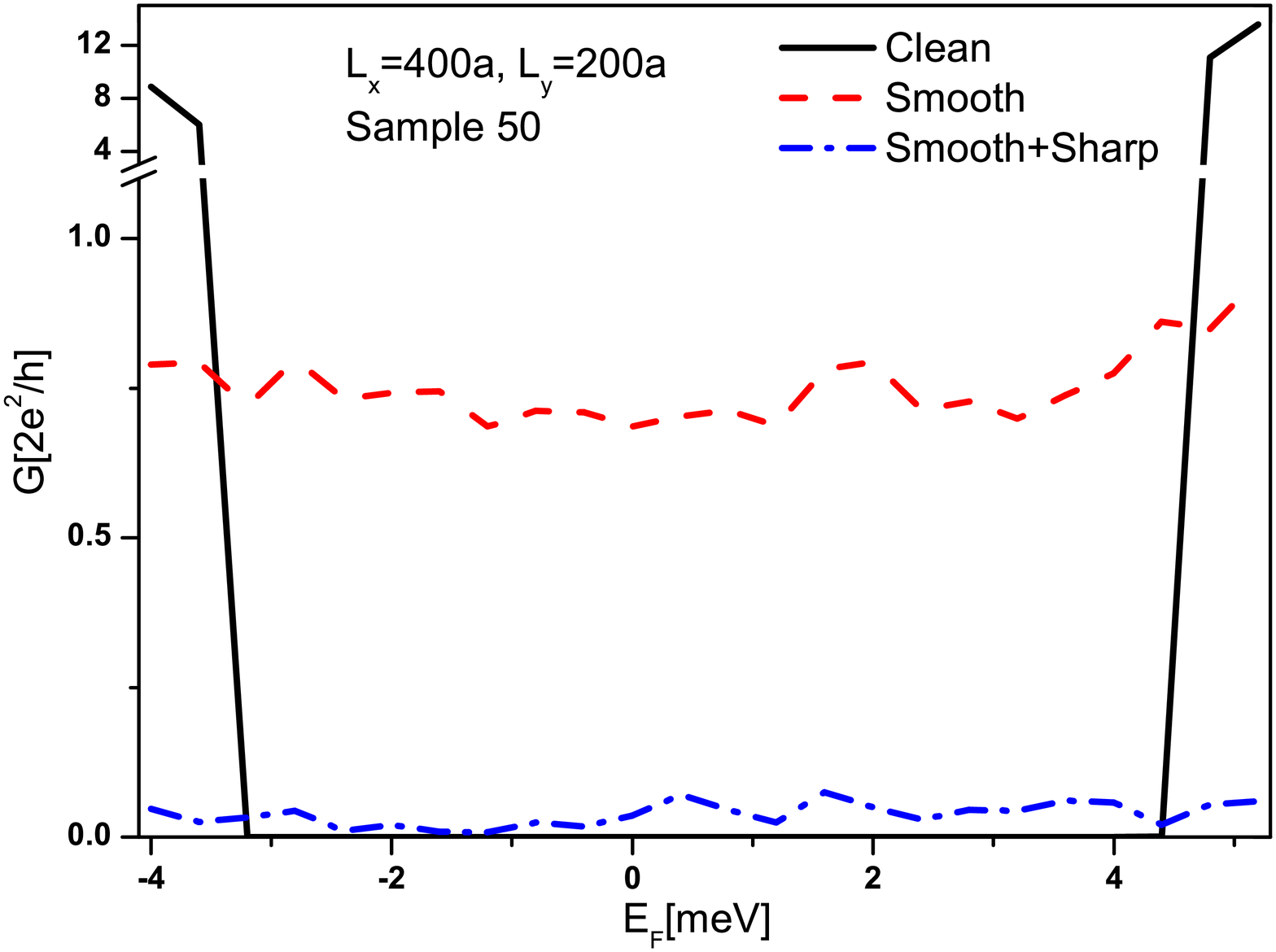}
\caption{(Color online) Bulk conductance in the presence of smooth disorder potential and smooth + sharp disorder potential.
In-gap ``extended" induced by smooth disorders will be localized by sharp disorders. The red dashed line is for the system
with only smooth disorder. The blue dashed line is for the system with both smooth and sharp dosorders.
}
\label{smooth_sharp}
\end{figure}

As shown in Fig.\ref{smooth_sharp}, smooth disorder potential will give rise to in-gap states with localization
length $\xi\gg Lx$, resulting in finite bulk conductance. These in-gap ``extended" states will then be further
localized by sharp impurity potential induced by dilute Si dopants, which significantly reduce the bulk conductance.

\end{document}